# An analysis of carrier dynamics in methylammonium lead triiodide perovskite solar cells using cross-correlation noise spectroscopy

Kevin Davenport[1], Fei Zhang[2], Mark Hayward[1], Logan Draper[1], Kai Zhu[2], Andrey Rogachev[1]

[1]Department of Physics and Astronomy, University of Utah, Salt Lake City, UT, USA.,
[2]National Renewable Energy Laboratory, Golden, CO, USA.

## Abstract

Using cross-correlation current noise spectroscopy, we have investigated carrier dynamics in methylammonium lead triiodide solar cells. This method provides a space selectivity for devices with planar multi-layered structure, effectively amplifying current noise contributions coming from the most resistive element of the stack. In the studied solar cells, we observe near full-scale shot noise, indicating the dominance of noise generation by a single source, likely the interface between the perovskite and the spiro-OMeTAD hole-transport layer. We argue that the strong $1/f$ noise term has contributions both from the perovskite layer and interfaces. It displays non-ideal dependence on photocurrent, $S \propto I^{1.4}$ (instead of usual $S \propto I^2$), which is likely due to current-induced halide migration. Finally, we observe generation-recombination noise. The relaxation time of this process grows linearly with photocurrent, which allows to attribute this contribution to bimolecular recombination in the perovskite bulk absorption layer. Extrapolating our results, we estimate that at the standard 1 sun illumination the electron-hole recombination time is 5 microseconds.

      To meet ever-increasing energy needs, a great deal of research has gone into finding cheap alternatives to silicon photovoltaics. One of the most promising materials are hybrid organic-inorganic perovskite solar cells (PSCs), which by now have reached certified power conversion efficiencies of more than 25% [1,2,3,4]. These materials have a tetragonal crystal structure at room temperature and chemical formula $ABX_3$. Here, A represents an organic cation, typically methylammonium ($CH_3NH_3^+$ or $MA^+$) or formamidinium ($H_2NCHNH_2^+$ or $FA^+$), B is a heavy metal such as $Pb^{2+}$ or $Sn^{2+}$, and X is a halide anion such as Cl, Br, or I. Perovskites are desirable due to low cost and ease of processing [3]. They offer a range of attractive qualities such as high carrier mobilities, long lifetimes and diffusion lengths [5], shallow intrinsic defect states [6], and low exciton binding energies [7].

      While much attention has been given to understanding the physics of the perovskite absorber layer, less is known about the dynamics at the interfaces between this layer and the electron and hole transport layers (ETL and HTL, respectively). One of the most commonly used HTL materials is the organic 2,2',7,7'-tetrakis(N,N-di-*p*-methoxyphenyl-amine)9,9'-spirobifluorene, also known as spiro-OMeTAD. Despite its popularity, there is still a limited understanding of its hole-transport properties. Further, spiro-OMeTAD films tend to be amorphous, leading to lower mobility, requiring doping [8] and making device reproducibility difficult. The material also suffers photoinduced alterations [9].

      The ETL is typically comprised of a transition metal oxide, with $TiO_2$ being the most widely used both as a contact layer as well as a mesoporous scaffolding for perovskite growth. This scaffolding is generally required as electron diffusion length in perovskites is much shorter than that of the holes [10]. While $TiO_2$ increases the device efficiency, it must be annealed at high temperature and introduces hysteresis and charge-trapping effects [11].

      In this paper, we use current noise spectroscopy to detect and characterize different relaxation processes observed in hybrid perovskite solar cells. We argue that current noise spectroscopy has a certain *spatial* selectivity, namely that it magnifies contributions from the most resistive elements of the stacked layers. In the case of hybrid perovskites, this appears to be the ETL/perovskite and perovskite/HTL interfaces.

      Noise spectroscopy analyzes fluctuations of a signal from its steady-state value. It has been used to characterize defects and relaxation processes in semiconducting devices [12,13,14], crystalline solar cells [15], light-induced defects in *a*-Si:H [16] as well as carrier kinetics [17] and metastable states [18] in perovskite solar cells. The technical advantage of our work comes in the use of the current noise cross-correlation technique [19], which provides several orders of magnitude better sensitivity and bandwidth than standard noise measurements. Using this technique, we have been able to resolve the frequency-independent shot noise contribution in fluorescent [20] and multilayered phosphoresecent [21] organic light emitting diodes (OLED).

The samples studied in the present paper, as shown in Fig. 1(a), consist of a 60-nm-thick compact TiO$_2$ layer grown on top of a fluorine-doped tin oxide (FTO)/glass substrate. The MAPbI$_3$ absorption layer was spin coated on top at two different thicknesses, $d$ = 200 nm (device S7) and 800 nm (device S12). This is topped, again by spin coating, with 180-200nm of spiro-OMeTAD followed by gold contacts. The active area of the devices is ~18 mm$^2$.

Solar cells were illuminated through the FTO/glass substrate using a narrow spectrum, high-power LED with a peak wavelength of 585-595nm, well above the MAPbI$_3$ bandgap of ~1.55 eV. The devices' $I(V)$ characteristics displayed a hysteresis, as shown in Fig. 1b. Previous studies suggest that this due to halide ion migration through the perovskite structure and (reversible) collection at the TiO$_2$/perovskite interface due to weak Ti-I-Pb bonds [22]. To avoid this issue, noise data were taken under short-circuit conditions and after "light-soaking" illumination for 30 minutes for each level of light intensity.

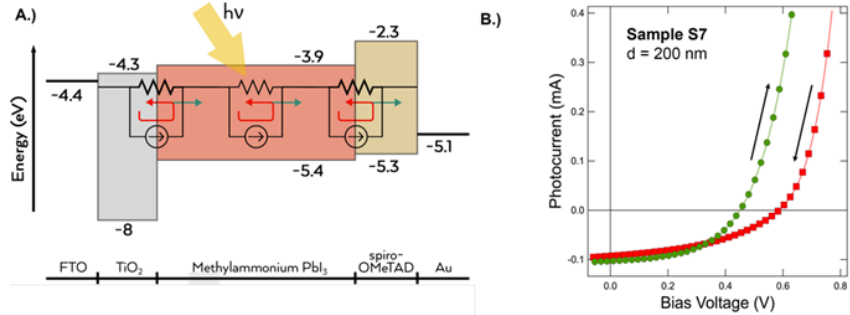

**Fig. 1.** Device structure and operation for the MAPbI$_3$ perovskite solar cells. (a) Device energy diagram; overlaid is equivalent circuit representing bulk and interfacial shot noise generators. (b) Typical $I(V)$ curve illustrating the observed device hysteresis. The arrows indicate the direction of the voltage sweep

Figure 2(a) shows a representative set of current noise spectral density curves taken from device S12 (MAPbI$_3$ thickness 800 nm) under different illuminations. The legend displays the measured short-circuit current, $I_{SC}$, for each curve. The black curves represent a fit to equation,

$$S_I = S_1 + \frac{S_2}{f^\alpha} + \mathrm{Re}\left[\frac{S_3}{1+(i2\pi f)^{1-b}}\right] + Af^2 \qquad (1)$$

where the first term is a frequency-independent term, the second represents 1/$f$ flicker noise, and the third is a generalized generation-recombination (GR) noise allowing for the dispersion of the relaxation time $\tau$. Note that when $b$ = 0, the GR term reverts to the familiar Lorentzian profile. The last term, $Af^2$, describes the residual background noise caused by capacitance of the devices as discussed in details in our previous works [20,21]. At the lowest illuminations, the GR is not detectable and $S_3$ = 0. With increasing illumination, however, a GR term emerges and is required for an accurate fit. This is illustrated by the dashed curve in Fig. 2(a) where all other terms are identical to the solid fit except that $S_3$ = 0, highlighting the need for the GR term in an accurate fit.

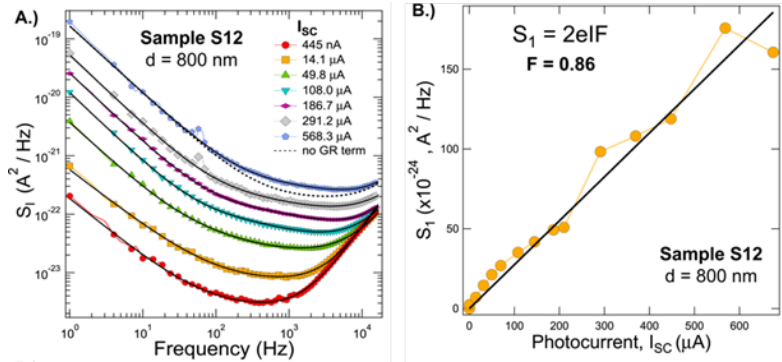

**Fig. 2.** Current noise data for S12 sample (MAPbI$_3$ thickness $d$ = 800 nm) (a) Current noise spectral density versus frequency. Solid lines are fits to the Eq. 1. The dashed black line shows a fit to Eq. 1 where $S_3$ = 0, illustrating the existence of GR noise. (b) Frequency-independent noise term $S_1$ versus photocurrent; linear fit, shown as a solid black line, gives Fano factor $F = 0.86$.

A plot of the frequency-independent term, $S_1$, versus photocurrent is shown in Fig. 2(b). The trend is clearly linear, a hallmark of shot noise; fit to the dependence $S_1 = 2eFI$ returns Fano factor $F = 0.86$. We estimated the possible thermal noise contribution using equation $S_{Th} = 4kTG$ and values for an impedance $G$ measured under short-circuit conditions [23]. This shows that $S_{Th}$ is at least an order of magnitude smaller than the observed noise and thus can be neglected.

The noise spectra for thinner S7 solar cell are shown in Fig. 3a. Compared and contrasted to the thicker S12 solar cell, no GR feature is resolved. Even when the GR magnitude set to zero, Eq. 1 provides good fit to the spectrum (solid lines in Fig. 3a). We believe, however, the distinction between two samples is apparent and reflects an inability of our fitting routine to detect small contributions, rather than an appearance of a new physical process in S12. The simplest explanation is that GR term is also present in S7 and, while unresolved on its own, provides a boost to the Fano factor, shifting it to unphysical super-Poissonian ($F > 1.1$) value (Fig. 3b).

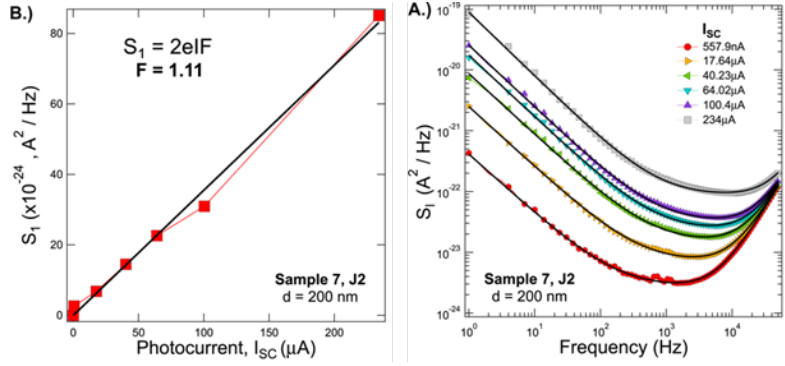

**Fig. 3.** Current noise data for sample S7 (MAPbI$_3$ thickness $d$ = 200 nm) (a) Noise spectral density versus frequency for indicated photocurrents. The black curves are fits to Eq.1 with GR magnitude set to zero (b) Frequency-independent noise extracted from the fits in (a) yielding a Fano factor of $F$=1.1.

The appearance of $F < 1$ in the thick device is more interesting. To understand its origin, let us look back at the device sequence shown in Fig. 1. To first approximation, the total noise can be represented by a series of noise sources, $i_n$, each self-shorted by its own internal resistance, $R_n$. Here sub-index $n$ indicates distinct elements of the stack and the interfaces between them. From Kirchhoff's law, the total noise current seen at the contacts is

$$I_T = \sum_n \frac{i_n R_n}{R_T} \qquad (2)$$

where $R_T$ is the total device resistance. If we assume that all noise generators $i_n$ are uncorrelated shot noise sources, the total current noise seen at the electrodes is

$$S_T = \langle I_T^2 \rangle = \sum_n \langle i_n^2 \rangle \left(\frac{R_n}{R_T}\right)^2 = 2eI \sum_n \left(\frac{R_n}{R_T}\right)^2 = 2eIF \qquad (3)$$

where the Fano factor, $F$, is a suppression of the overall shot noise. The sum in Eq. 3 is dominated by the most resistive term; this feature for multi-layered devices provides a space selectivity to the noise signals.

Shot noise reflects the discreet nature of electric charge and typically appears due to the random transfer of electrons (or holes) across an energy barrier. In our experiment, the close-to-unity value of the Fano factor indicates the dominant contribution of a single resistive element in stack. Given the abrupt interface between the perovskite and the spiro-OMeTAD hole transport layer, as compared to the diffused scaffolding of the TiO$_2$, it is likely that the shot noise is dominated by a Schottky barrier at the perovskite/HTL interface.

Let us now discuss the other parameters extracted from the noise spectra fits. Fig. 4a shows the $1/f$ noise magnitude in log-log scale, for S7 in red (circles) and S12 in yellow (squares). The magnitude of the noise in thin device S7 is roughly twice larger than in S12. Assuming that the properties of interfaces in both devices are the same, the larger noise in S7 should come from the enhanced $1/f$ noise in the bulk perovskite layer. This is an expected tendency. Indeed, in inorganic semiconductors, flicker noise is predicted to follow the dependence $\langle S_{1/f} \rangle \propto I^2 / V_3$ [24], that is that a sample with a smaller volume $V_3$ generates larger noise signal. This holds for

$1/f$ noise generated by either mobility or concentration fluctuations [24]. In analysis of $1/f$ noise, we can still use the reasoning based on Eq. 3. However, unlike the shot noise, $1/f$ noise has no universal origin. The elements of the stack can potentially have very different $1/f$ noise terms, perhaps by orders of magnitude. The very fact that we observe dependence of noise on perovskite thickness suggests the $1/f$ noise generator for bulk perovskite layer has much larger magnitude than $1/f$ noise coming from interfaces.

We further notice that $1/f$ noise grows with current as $S_{1/f} = AI^{\alpha}$ with an exponent $\alpha \approx 1.5$. The standard value $\alpha = 2$ corresponds to the processes caused by traps whose concentration and properties do not depend on driven current. Deviation from $\alpha = 2$ suggests the presence of traps and defects induced by current or light. Our results suggest that the induced defects are present both in the bulk perovskite layer and at interfaces. The likely source of these current-driven defects is the migration and interfacial collection of the halide ions, though its exact origin cannot be established solely from noise measurements.

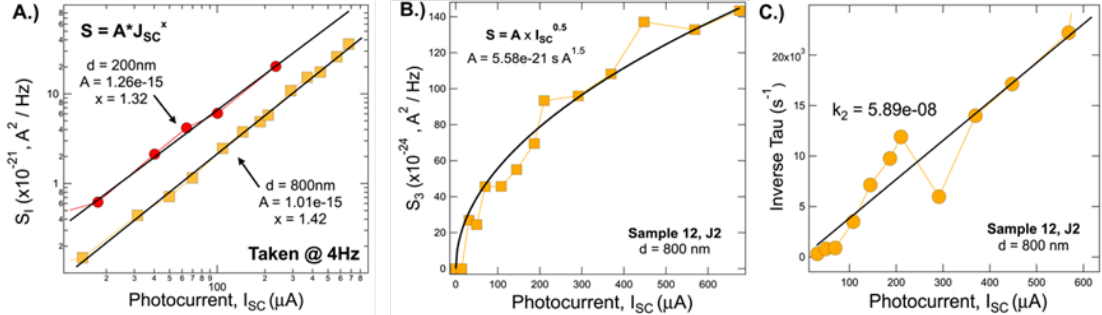

**Fig. 4.** Noise parameters extracted from curve fits. (a) Flicker noise, sampled at 4 Hz, in the thin device S7 (shown in red circles) and the thick device S12 (shown in yellow squares) as a function of photocurrent. Black lines represent a fit to a generic power law. (b) The GR noise magnitude as a function of photocurrent, showing a square root dependence. (c) Extracted GR timescale as a function of photocurrent.

Let us now discuss the generation-recombination term (third term in Eq. 1). Figure 4(b) shows its magnitude as a function of photocurrent, which appears to follow a square-root dependence. The relaxation exponent, $b$, was found to be 0.32, indicating a moderate dispersion of the relaxation time. Most importantly, the relationship between the inverse relaxation time and the photocurrent was found to be linear, as shown in Fig 4(c).

Phenomenologically, the rate of a recombination process in semiconductor depends on number of carriers involved in the elementary recombination act and, in the simplest case when electrons and holes have the same concentration $n$, can be written as [25,26],

$$R(n) = k_1 n + k_2 n^2 + k_3 n^3. \qquad (5)$$

The first term represents monomolecular recombination, in which carriers recombine via an intermediate process of a capture by a trap. The second is bimolecular recombination involving two carriers such as direct band-to-band recombination of electron and hole. Finally, there is trimolecular recombination in which three carriers are involved; the canonical example of this is Auger recombination where an electron and hole recombine with an energy transfer to a third particle.

The noise spectroscopy measures relaxation time of small fluctuations from a steady state value $n_0$. So the relaxation time determined by this method is

$$\tau = \left(\frac{\partial R(n)}{\partial n}\right)^{-1}_{n_0}. \qquad (6)$$

Under steady state operation the photocurrent generated by a solar cell is proportional to the steady state, non-equilibrium concentration of carriers. Hence, the experimentally observed linear dependence of $1/\tau$ on current allows us to identify the microscopic process as bimolecular recombination.

$$\frac{1}{\tau} = 2k_2 n_0 \propto I_{SC} \qquad (7)$$

To compare our results quantitatively with values of the recombination time obtained by other methods, we need to extend it to the standard condition of 1 sun, since $\tau$ depends on light intensity. For the S12 device, the highest intensity present in Fig. 4(c) corresponds to $568\ \mu A$ or to 3.1 mA/cm$^2$. Under the standard AM 1.5 conditions, our devices exhibit a typical current density of 20-25 mA/cm$^2$. Assuming that the recombination rate continues to grow linearly with photocarrent, we estimate $\tau_{1sun} \approx 5$ μs. This number has the same order of magnitude as the lifetime reported for a series of MAPb(I$_{1-x}$Br$_x$)$_3$ perovskites, $\tau_{1sun} \approx 0.4-2$ μs, under 1 sun illumination [27] and is roughly consistent with the conclusion made in Ref. [28] that under ambient sunlight the electron-hole recombination is very slow, in timescale of tens of microseconds.

While the presense of monomolecular and Auger processes is not directly observed in the ranges of illumination studied, this does not mean that they are completely absent from the device. Monomolecular recombination is known to occur via the Shockley-Reed-Hall mechanism at interfacial trap sites. Due to the energetic distribution of these sites, however, it is likely that they would contribute to 1/*f* noise. Auger recombination is known to be unexpectedly strong in halide perovskites [29], though only at very high illumination. This has little bearing on typical operation, though it is important in solar concentrators and of perovskite-based LEDs, where the charge carrier concentrations are orders of magnitude above solar cells in full sun [30].

In summary, through noise spectroscopy, we observe a rich set of data highlighting important dynamics inside a modern MAPbI$_3$ photovoltaic cell. We see shot noise with near-unity Fano factor in devices with 4-fold difference in thickness. This indicates that a single interface or step is the dominating shot noise generator, most probably the perovskite/spiro-OMeTAD interface. The *1/f* noise has contributions both from the light-absorbing perovskite layer and interfaces. The dependence of *1/f* noise on photocurrent indicates presence of defects induced by light and/or current. In the thicker device, we observe a GR contribution, which we attribute to bimolecular recombination occurring in the bulk perovskite. The extracted relaxation time agrees in order of magnitude with the values reported in literature. In the analysis of the noise data, we used the important advantage of the current noise spectroscopy, namely spatial selectivity in which the most resistive elements in the device dominate the noise profile, a method applicable to other devices with a complex stacked structure. Overall, we see that current noise spectroscopy is a useful and inexpensive method for material and device characterization.

**Acknowledgments**
KD and AR gratefully acknowledge that this work has been supported by NSF grants DMR1611421 and DMR1904221. The work at the National Renewable Energy Laboratory (NREL) was supported by De-risking Halide Perovskite Solar Cells program of the National Center for Photovoltaics, funded by Office of Energy Efficiency and Renewable Energy, Solar Energy Technologies Office, U.S. Department of Energy under Contract No. DE-AC36-08GO28308 with Alliance for Sustainable Energy, Limited Liability Company (LLC), the Manager and Operator of NREL. The views expressed in the article do not necessarily represent the views of the DOE or the U.S. Government. The U.S. Government retains and the publisher, by accepting the article for publication, acknowledges that the U.S. Government retains a nonexclusive, paid-up, irrevocable, worldwide license to publish or reproduce the published form of this work, or allow others to do so, for U.S. Government purposes.